\documentclass[11pt]{article}
\usepackage[margin=1in]{geometry}
\usepackage{amsmath,amssymb,bm}
\usepackage{graphicx}
\usepackage{float}
\usepackage[section]{placeins}
\usepackage{booktabs}
\usepackage{siunitx}
\usepackage[backend=biber,style=numeric,sorting=none,natbib=true]{biblatex}
\usepackage[colorlinks=true,linkcolor=blue,citecolor=blue,urlcolor=blue]{hyperref}
\usepackage{caption}
\usepackage{subcaption}
\usepackage{microtype}
\makeatletter
\let\ps@plain\ps@empty
\makeatother
\title{Sectorial customized corneal crosslinking for keratoconus: an inverse biomechanical design study with an anisotropic reduced shell finite-element surrogate}
\author{Juan Sumaya-Mart\'inez\thanks{Corresponding author: j.sumaya2011@gmail.com}\\
Faculty of Sciences, Universidad Aut\'onoma del Estado de M\'exico, Toluca 50000, M\'exico
\and A. Altamirano-Torres\\
Centro de Microcirug\'ia Ocular, Atlacomulco, Estado de M\'exico, M\'exico}
\date{}

\begin{document}
\pagestyle{empty}
\maketitle
\thispagestyle{empty}

\begin{abstract}
\textbf{Purpose.} Keratoconus is a localized biomechanical failure of the cornea that appears clinically as thinning, cone formation and higher-order optical aberrations. Standard corneal collagen crosslinking (CXL) is effective for stabilization, but it is usually applied over a broad central area. Because many cones are decentered, this broad stiffening pattern may not be the most efficient way to reduce coma or to redistribute stromal stress. We frame customized CXL as an inverse biomechanical--optical design problem.\\
\textbf{Methods.} We built a patient-inspired, anisotropic reduced shell finite-element surrogate of an inferotemporal keratoconus-like cornea. Disease was introduced through local thinning and local loss of effective stiffness. Crosslinking was represented as a spatial stiffness-modulation field that could, in principle, arise from a diffusion--reaction process such as genipin delivery. The mesh contained 1090 nodes and 2091 triangular elements. The numerical eye had an 11.5-mm diameter, a 7.80-mm anterior radius, a 520-$\mu$m central stromal thickness, a cone centered at $(x,y)=(0.35,-1.15)$ mm, a 64\% local stiffness reduction, a 125-$\mu$m thinning field and a 15-mmHg posterior pressure load. Uniform, cone-sector, partial-annular, coma-gradient, cone-Gaussian and inverse-smooth masks were compared using pressure displacement, a strain-energy concentration proxy, a Kmax-equivalent severity index and Zernike metrics over a 6-mm pupil.\\
\textbf{Results.} In the untreated model, cone displacement was 280.2 $\mu$m, vertical coma was 13.41 $\mu$m and higher-order RMS was 6.57 $\mu$m. Uniform stiffening reduced cone displacement to 142.3 $\mu$m and HOA RMS to 3.21 $\mu$m, but residual vertical coma remained 6.25 $\mu$m. Cone-sector and coma-gradient masks reduced coma more strongly, to 4.77 and 4.68 $\mu$m, respectively, although with different mechanical trade-offs. The inverse-smooth mask gave the most balanced response: cone displacement 154.6 $\mu$m, vertical coma 5.22 $\mu$m, HOA RMS 2.96 $\mu$m and Kmax-equivalent severity 48.81 D versus 52.50 D in the untreated case.\\
\textbf{Conclusions.} Sectorial customized CXL should not be viewed simply as stiffening the steepest part of the cornea. It is better understood as spatial control of a weakened, anisotropic shell. The simulations suggest that localization can improve coma targeting, but abrupt masks may introduce trade-offs. Smooth inverse-designed masks offer a more conservative design principle. Genipin-mediated stiffening is used here as a modeling platform, not as a validated clinical protocol.
\end{abstract}


\section{Introduction}
Keratoconus is often introduced as an optical disease: the cornea steepens, irregular astigmatism increases and visual quality declines. That description is clinically useful, but it hides the mechanical origin of the problem. The cornea is a hydrated, prestressed and anisotropic shell. It is reinforced by collagen lamellae, supported at the limbus and continuously loaded by intraocular pressure. From this viewpoint, the optical phenotype of keratoconus is the visible result of a localized loss of load-bearing capacity.

Riboflavin/ultraviolet-A corneal collagen crosslinking is the standard conservative treatment for progressive keratoconus \citep{Spoerl1998,Wollensak2003,Hersh2017,Raiskup2015}. Its main clinical purpose is to arrest progression. Visual improvement may occur, but conventional CXL was not designed primarily as a shape-optimization procedure. This distinction matters because keratoconus is rarely symmetric. The cone is often inferior or inferotemporal, stromal thinning is localized, posterior elevation is asymmetric and epithelial remodeling can mask part of the stromal deformation. A spatially heterogeneous disease therefore receives, in many protocols, a comparatively broad and uniform stiffening field.

Customized, topography-guided and localized CXL address this mismatch by varying the treatment pattern over the cornea \citep{Cassagne2017,VorobichikBerar2025,Frigelli2025}. These approaches are clinically attractive because they try to concentrate the stiffening effect where the ectatic process is most relevant, while avoiding unnecessary treatment of relatively normal tissue. However, topography alone does not tell us why a region is steep. Local curvature can reflect stromal weakening, thinning, posterior displacement, epithelial compensation, altered boundary conditions or several of these factors acting together. For that reason, the treatment-planning question should not be limited to ``where is the cornea steepest?'' A more useful question is: where, and with what spatial profile, should stiffness be increased to improve both mechanical stability and optical quality?

Finite-element modeling is well suited to this question. Previous work by Pandolfi and collaborators has shown that corneal behavior depends strongly on geometry, pressure, anisotropy and regional material properties \citep{PandolfiManganiello2006,Sanchez2014,Pandolfi2020}. Broader reviews of corneal FEM have also emphasized that geometry, constitutive assumptions, boundary conditions and material calibration can strongly affect predictions \citep{Pang2024}. More recent microstructural and discrete-to-continuum models further support the idea that keratoconus-like deformation and thinning are tied to collagen architecture and crosslink degradation \citep{Gizzi2021,DeBellisPandolfi2024,Kory2024}. These studies motivate a treatment-design framework in which CXL is modeled not as a geometric correction imposed on the surface, but as a spatially controlled change in stiffness.

These modeling studies set a high standard for any computational CXL proposal: the model should state what is patient-specific, what is phenomenological, and what is only a reduced surrogate for the true three-dimensional stromal mechanics. Here we formulate customized CXL as an inverse biomechanical--optical design problem. The aim is not to propose genipin for immediate human treatment. Instead, genipin-mediated crosslinking is used as a model biochemical platform because its delivery can be represented by diffusion--reaction and its mechanical action can be encoded as local stiffness modulation \citep{Gharaibeh2018,Tang2019}. The purpose of the study is to compare how uniform, sectorial, annular, gradient and inverse-designed stiffening fields redistribute deformation and alter optical readout in a reproducible keratoconus-like model.

\section{Conceptual framework}
\subsection{Keratoconus as localized loss of stromal stability}
Let the effective pre-treatment stromal stiffness be written as
\begin{equation}
E_{\mathrm{KC}}(r,\theta,z)=E_{\mathrm{H}}(r,\theta,z)\left[1-\delta_{\mathrm{KC}}(r,\theta,z)\right],
\end{equation}
where $E_{\mathrm{H}}$ denotes a reference healthy stiffness field and $\delta_{\mathrm{KC}}$ is a localized weakening field. A minimal decentered-cone representation is
\begin{equation}
\delta_{\mathrm{KC}}(r,\theta,z)=\delta_0 \exp\left[-\frac{(r-r_c)^2}{2\sigma_r^2}-\frac{\operatorname{ang}(\theta-\theta_c)^2}{2\sigma_\theta^2}\right] f_z(z),
\end{equation}
where $(r_c,\theta_c)$ locates the ectatic zone, $\sigma_r$ and $\sigma_\theta$ set its spread, $f_z$ describes depth dependence and $\operatorname{ang}(\cdot)$ wraps the angular difference to $[-\pi,\pi]$.

After crosslinking, the local stiffness is represented as
\begin{equation}
E_{\mathrm{post}}(r,\theta,z)=E_{\mathrm{KC}}(r,\theta,z)\left[1+\eta(r,\theta,z)\right],
\end{equation}
where $\eta$ is the relative stiffening. Uniform CXL corresponds to a broad radially symmetric $\eta$. Customized CXL corresponds to an angularly localized or smoothly graded $\eta$. Inverse-designed CXL seeks the admissible $\eta$ that minimizes a biomechanical--optical objective while respecting dose and safety constraints.

\subsection{Genipin as a model stiffness-modulation agent}
For a chemical crosslinking agent such as genipin, stromal delivery can be approximated by
\begin{equation}
\frac{\partial C}{\partial t}=D\nabla^2 C-k_r C+S(r,\theta,z,t),
\end{equation}
where $C$ is concentration, $D$ is effective diffusivity, $k_r$ is a reaction or consumption rate and $S$ is the spatial application source. A patterned topical exposure can be written as
\begin{equation}
S(r,\theta,z,t)=S_0 A(r,\theta)g(z)h(t),
\end{equation}
where $A(r,\theta)$ is the treatment mask, $g(z)$ accounts for epithelial or boundary transport and $h(t)$ defines exposure duration.

The cumulative exposure is
\begin{equation}
\Gamma(r,\theta,z)=\int_0^T C(r,\theta,z,t)\,dt,
\end{equation}
and the local stiffness increment is represented by a saturating law
\begin{equation}
\eta(r,\theta,z)=\eta_{\max}\frac{\Gamma(r,\theta,z)^m}{\Gamma_{50}^m+\Gamma(r,\theta,z)^m}.
\end{equation}
This phenomenological law is not intended to replace biochemical calibration. It provides a way to map a spatial drug-delivery field into a mechanical stiffening field.

\section{Coupled biomechanical--optical model}
\label{sec:coupled_model}
The model links geometry, local mechanical damage, treatment-induced stiffening, finite-element pressure response and optical readout in one pipeline. We deliberately describe it as a \emph{patient-inspired anisotropic reduced shell finite-element surrogate}. Its role is to compare treatment patterns and to test inverse-design logic. It is not intended to predict the postoperative shape of a specific clinical eye. In a future translational workflow, the effective stiffness field introduced below would need to be calibrated against independent biomechanical measurements, for example Brillouin microscopy, optical coherence elastography, inflation testing or inverse FEM \citep{Scarcelli2013,Shao2019,Zhao2024,Hammoud2025,Vinciguerra2025}.

\subsection{Reference geometry and keratoconus-like thinning}
The cornea is represented over a circular domain
\begin{equation}
\Omega=\{(x,y):x^2+y^2\leq R_c^2\},
\end{equation}
where $R_c=5.75$ mm in the numerical study. The anterior and posterior surfaces are denoted by $z_a(x,y)$ and $z_p(x,y)$, with local stromal thickness
\begin{equation}
h(x,y)=z_a(x,y)-z_p(x,y).
\end{equation}
The reference anterior surface is taken as a spherical cap
\begin{equation}
z_a^0(x,y)=R_a-\sqrt{R_a^2-x^2-y^2},
\end{equation}
with anterior radius $R_a=7.80$ mm. A decentered keratoconus-like defect is introduced by a Gaussian field
\begin{equation}
G_c(x,y)=\exp\left[-\frac{(x-x_c)^2}{2\sigma_x^2}-\frac{(y-y_c)^2}{2\sigma_y^2}\right],
\end{equation}
where $(x_c,y_c)$ is the cone center. The diseased thickness map is
\begin{equation}
h_{\rm KC}(x,y)=h_0(x,y)-\Delta h_{\max}G_c(x,y),
\end{equation}
so that local thinning, rather than anterior shape alone, contributes to the ectatic mechanical response. This step is important: a cone-like surface alone would be a geometric perturbation, whereas keratoconus is a coupled change in shape, thickness and load-bearing capacity.

\subsection{Mechanical damage field}
The diseased effective modulus is written as
\begin{equation}
E_{\rm KC}(x,y,z)=E_H(x,y,z)\left[1-\delta_{\rm KC}(x,y,z)\right],
\end{equation}
with
\begin{equation}
\delta_{\rm KC}(x,y,z)=\delta_0 G_c(x,y)f_z(z).
\end{equation}
Here $E_H$ is the reference healthy modulus, $\delta_0$ is the maximum local softening and $f_z(z)$ is a depth-weighting function. In the reduced two-dimensional shell implementation, depth dependence is collapsed into an effective thickness-weighted modulus,
\begin{equation}
\bar E_{\rm KC}(x,y)=\frac{1}{h_{\rm KC}(x,y)}\int_{z_p}^{z_a} E_{\rm KC}(x,y,z)\,dz.
\end{equation}
This averaging is one of the main approximations of the surrogate model. A future patient-specific implementation should replace it with a three-dimensional hyperelastic formulation with explicit stromal depth dependence.

\subsection{Crosslinking-induced stiffness field}
Crosslinking is modeled as a local multiplicative increase in the diseased stiffness,
\begin{equation}
E_{\rm post}(x,y,z)=E_{\rm KC}(x,y,z)\left[1+\eta(x,y,z)\right],
\label{eq:post_stiffness_v3}
\end{equation}
where $\eta$ is the relative stiffening produced by the treatment. In the reduced simulation this becomes
\begin{equation}
\bar E_{\rm post}(x,y)=\bar E_{\rm KC}(x,y)\left[1+\bar\eta(x,y)\right].
\end{equation}
The field $\eta$ may be generated by riboflavin--UVA fluence, oxygen-modulated photochemistry, genipin delivery or another crosslinking strategy. In this manuscript, genipin is used only as a model chemical agent because its delivery can be naturally represented by diffusion--reaction and a saturating dose--response law.

\subsection{Treatment-mask basis}
The two-dimensional treatment mask $A(x,y)$ determines where crosslinking is applied. The simple masks used in the study are:
\begin{align}
A_{\rm uni}(r)&=H(R_{\rm CXL}-r),\\
A_{\rm sec}(r,\theta)&=H(r-r_1)H(r_2-r)H(\Delta\theta-|\operatorname{ang}(\theta-\theta_c)|),\\
A_{\rm ann}(r,\theta)&=H(r-r_{\rm in})H(r_{\rm out}-r)H(\Delta\theta-|\operatorname{ang}(\theta-\theta_c)|),\\
A_{\rm coma}(r,\theta)&=A_0(r)\left[1+\alpha\cos(\theta-\theta_c)\right],\\
A_{\rm gauss}(x,y)&=G_c(x,y).
\end{align}
The inverse-designed mask is written as a bounded smooth combination
\begin{equation}
A_{\rm inv}(x,y)=\mathcal P_{[0,1]}\left[\sum_{j=1}^{N_b} c_j A_j(x,y)\right],
\end{equation}
where $\mathcal P_{[0,1]}$ clips the mask to the admissible dose interval and the coefficients $c_j$ are selected by minimizing the biomechanical--optical objective. Sharp masks are allowed as test cases, but the objective explicitly penalizes excessive dose and steep spatial gradients because abrupt stiffness transitions may introduce stress concentrations.

\subsection{Reduced finite-element equilibrium problem}
The unknown $w(x,y)$ represents the dominant pressure-induced anterior displacement mode. The reduced shell stiffness entering the weak form is
\begin{equation}
T(x,y)=\frac{\bar E_{\rm post}(x,y)h_{\rm KC}(x,y)}{1-\nu^2},
\end{equation}
for the isotropic limit. In the enhanced anisotropic implementation, $T\mathbf I$ is replaced by the shell-stiffness tensor
\begin{equation}
\mathbf D_s(x,y)=T(x,y)\left[(1-f_a)\mathbf I+2f_a\left(0.65\,\mathbf e_\theta\otimes\mathbf e_\theta+0.35\,\mathbf e_r\otimes\mathbf e_r\right)\right],
\label{eq:anisotropic_tensor_v3}
\end{equation}
where $f_a$ is a phenomenological fiber-reinforcement fraction and $(\mathbf e_r,\mathbf e_\theta)$ are radial and circumferential unit vectors. The weak form is
\begin{equation}
\int_\Omega \nabla v^T\mathbf D_s(x,y)\nabla w\,d\Omega+
\int_\Omega k_s(x,y)vw\,d\Omega=
\int_\Omega q(x,y)v\,d\Omega,
\label{eq:weak_form_v3}
\end{equation}
for all test functions $v$. The term $k_s(x,y)$ is a weak stromal/scleral support and $q(x,y)$ is the pressure-derived load. In the keratoconus-like region the load is amplified and the support is reduced to mimic the increased pressure susceptibility of a thinned and softened ectatic shell,
\begin{equation}
q(x,y)=q_0\left[1+\beta_q G_c(x,y)\right]\left(\frac{h_0}{h_{\rm KC}(x,y)}\right)^2.
\end{equation}
The limbal rim is fixed. This boundary condition is sufficient for comparing treatment masks under identical assumptions, although a scleral shell or elastic limbal support would be preferable for patient-specific prediction.

\subsection{Mechanical energy and cone metrics}
The local strain-energy proxy is
\begin{equation}
\Psi(x,y)=\frac{1}{2}\nabla w^T\mathbf D_s(x,y)\nabla w.
\end{equation}
A cone-weighted concentration metric is then defined by
\begin{equation}
\Psi_{\rm cone}=\frac{\int_\Omega \Psi(x,y)G_c(x,y)\,d\Omega}{\int_\Omega G_c(x,y)\,d\Omega}.
\end{equation}
The mean cone displacement is similarly
\begin{equation}
d_{\rm cone}=\frac{\int_\Omega w(x,y)G_c(x,y)\,d\Omega}{\int_\Omega G_c(x,y)\,d\Omega}.
\end{equation}
These quantities are used to distinguish global flattening from actual reduction of the ectatic mechanical response.

\subsection{Optical reconstruction}
The deformed anterior surface is reconstructed as
\begin{equation}
z_a^{\rm def}(x,y)=z_a^{\rm KC}(x,y)+w(x,y).
\end{equation}
A curvature-derived power proxy is obtained from the mean curvature of $z_a^{\rm def}$,
\begin{equation}
P(x,y)\simeq (n_c-n_a)\,\kappa(x,y),
\qquad
\kappa=\nabla\cdot\left(\frac{\nabla z_a^{\rm def}}{\sqrt{1+|\nabla z_a^{\rm def}|^2}}\right).
\end{equation}
Because the reduced model does not include full anterior--posterior ray tracing, this quantity is reported as a Kmax-equivalent severity index, not as clinical keratometry.

The anterior-surface wavefront is approximated by
\begin{equation}
W(\rho,\phi)=(n_c-n_a)\left[z_a^{\rm def}(\rho,\phi)-z_{\rm ref}(\rho,\phi)\right],
\end{equation}
and expanded in Zernike polynomials,
\begin{equation}
W(\rho,\phi)=\sum_{n,m}a_n^m Z_n^m(\rho,\phi).
\end{equation}
The main optical quantities are vertical coma $Z_3^{-1}$, horizontal coma $Z_3^{1}$, spherical aberration $Z_4^0$ and higher-order RMS after removing piston, tilt, defocus and astigmatism.

\subsection{Inverse-design objective and safety constraints}
The treatment is selected by minimizing
\begin{align}
\mathcal J=&\;w_K\left(\frac{K_{\rm eq}-K_{\rm targ}}{K_{\rm eq,0}}\right)^2
+w_C\left(\frac{Z_3^{-1}}{Z_{3,0}^{-1}}\right)^2
+w_D\left(\frac{d_{\rm cone}}{d_{{\rm cone},0}}\right)^2 \\
&+w_E\left(\frac{\Psi_{\rm cone}}{\Psi_{{\rm cone},0}}\right)^2
+w_R\int_\Omega |\nabla\bar\eta|^2\,d\Omega
+w_Q\int_\Omega \bar\eta^2\,d\Omega.
\label{eq:objective_v3}
\end{align}
The last two terms penalize sharp masks and excessive dose. The admissible set includes
\begin{equation}
0\leq \eta\leq\eta_{\max},\qquad h(x,y)\geq h_{\min},\qquad \eta(x,y,z_{\rm endo})\approx 0,
\end{equation}
together with practical constraints on endothelial exposure, dose smoothness and maximum allowed stiffness gradient. This formulation keeps the optimization clinically interpretable. The goal is not simply to stiffen the cone, but to balance stabilization, optical improvement, dose economy and safety.

\section{Treatment masks and inverse design}
\begin{figure}[t]
\centering
\includegraphics[width=0.98\linewidth]{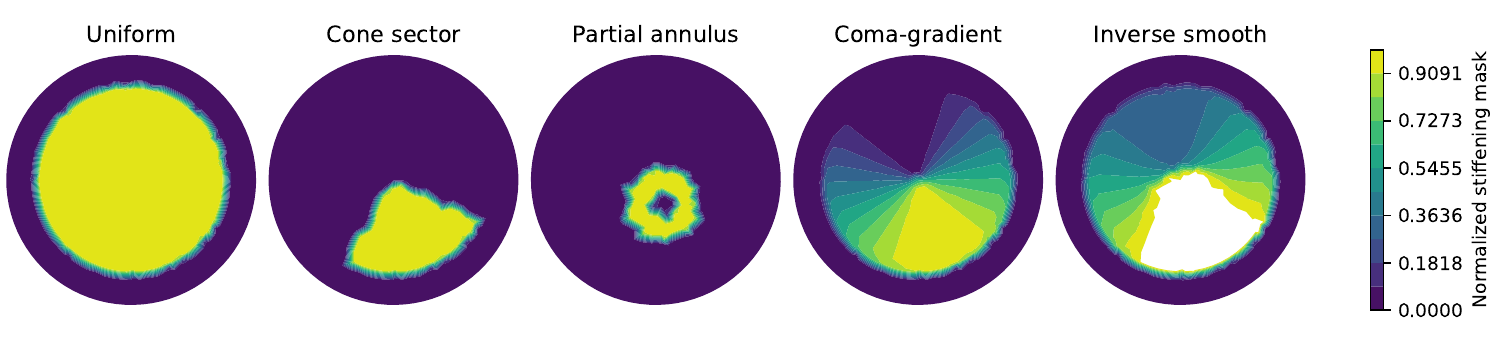}
\caption{Treatment masks used in the enhanced simulation. The inverse-smooth mask is a bounded smooth combination of uniform, cone-sector, partial-annular, coma-gradient, counter-gradient and cone-Gaussian bases.}
\label{fig:masks}
\end{figure}

Using the coupled model in Sec.~\ref{sec:coupled_model}, we compared five explicit treatment bases and one inverse-composite mask (Fig.~\ref{fig:masks}). The uniform mask represents broad central stiffening. The cone-sector mask targets the decentered ectatic direction. The partial-annular mask reinforces tissue around the cone while reducing dose at the thinnest point. The coma-gradient mask introduces a smooth angular correction aligned with the dominant coma direction. The cone-Gaussian mask tests the effect of stiffening the cone center. The inverse-smooth mask combines these bases under bounded dose and smoothness constraints.

The inverse problem was written as
\begin{equation}
\eta^*(r,\theta,z)=\arg\min_{\eta\in \mathcal A}\mathcal J[\eta],
\end{equation}
with admissible set $\mathcal A$ enforcing non-negative stiffening, bounded dose, smoothness and endothelial-safety restrictions. The reduced objective used in this study was
\begin{align}
\mathcal J=&\;w_K\left(\frac{K_{\mathrm{eq}}}{K_{\mathrm{eq},0}}\right)^2+w_C\left(\frac{|Z_3^{-1}|}{|Z_{3,0}^{-1}|}\right)^2+w_R\left(\frac{\mathrm{RMS}_{\mathrm{HOA}}}{\mathrm{RMS}_{\mathrm{HOA},0}}\right)^2 \\
&+w_D\left(\frac{d_{\mathrm{cone}}}{d_{\mathrm{cone},0}}\right)^2+w_S\left(\frac{\Psi}{\Psi_0}\right)^2+w_\eta \langle \eta/\eta_{\max}\rangle^2.
\end{align}
Here $K_{\mathrm{eq}}$ is the Kmax-equivalent severity index defined below, $Z_3^{-1}$ is vertical coma, $d_{\mathrm{cone}}$ is mean pressure displacement in the cone region, $\Psi$ is a strain-energy concentration proxy and subscript 0 denotes the untreated model.

\section{Enhanced reduced finite-element formulation}
\subsection{Geometry, material field and loading}
The numerical model was designed to move beyond a purely conceptual comparison while remaining transparent and reproducible. It is still a reduced shell surrogate rather than a full three-dimensional hyperelastic cornea, but it includes local thinning, local loss of stromal/scleral support, a denser mesh and anisotropic collagen-like reinforcement in the weak form.

The computational domain was a circular corneal disk of radius 5.75 mm. The mesh contained 1090 nodes and 2091 triangular elements. The keratoconus-like defect was centered at $(0.35,-1.15)$ mm with Gaussian widths $\sigma_x=0.82$ mm and $\sigma_y=0.78$ mm. The central stromal thickness was 520 $\mu$m, and the local thinning field reduced thickness by 125 $\mu$m near the cone. The healthy effective modulus was set to 0.55 MPa, with a maximum local keratoconus-like softening of 64\%. The posterior pressure load was 15 mmHg.

\begin{table}[t]
\centering
\caption{Input parameters for the enhanced reduced finite-element simulation. These values define a reproducible numerical eye for treatment-pattern comparison; they are not universal clinical constants.}
\label{tab:inputs}
\begin{tabular}{lll}
\toprule
Parameter & Value & Role \\
\midrule
Computational corneal diameter & 11.5 mm & Circular FEM domain \\
Anterior radius of curvature & 7.80 mm & Reference spherical cap \\
Central stromal thickness & 520 $\mu$m & Baseline thickness \\
Minimum cone thickness & $\approx$395 $\mu$m & Localized thinning after disease field \\
Cone center & $(0.35,-1.15)$ mm & Inferotemporal decentration \\
Cone Gaussian width & $(0.82,0.78)$ mm & Softening/thinning spread \\
IOP & 15 mmHg = 0.0020 MPa & Posterior pressure load \\
Healthy effective modulus & 0.55 MPa & Reduced shell modulus \\
Maximum local softening & 64\% & Keratoconus-like stiffness loss \\
Poisson ratio & 0.49 & Nearly incompressible approximation \\
Anisotropic fiber fraction & 0.34 & Circumferential/radial collagen proxy \\
Maximum CXL stiffening & $\eta_{\max}=2.40$ & Up to 3.4$\times$ local $E_{\rm KC}$ \\
Pupil diameter & 6.0 mm & Zernike analysis \\
Mesh & 1090 nodes, 2091 triangles & Linear triangular finite elements \\
\bottomrule
\end{tabular}
\end{table}

\subsection{Weak form}
The scalar field $w(x,y)$ represents the dominant pressure-induced anterior displacement mode of the corneal shell. The finite-element equilibrium problem was assembled as
\begin{equation}
\int_\Omega \nabla v^T \mathbf D_s(x,y)\nabla w\,d\Omega + \int_\Omega k_s(x,y)vw\,d\Omega=\int_\Omega q(x,y)v\,d\Omega,
\end{equation}
where $v$ is the test function, $k_s$ is a weak stromal/scleral support term and $q(x,y)$ is the pressure-derived load. The anisotropic shell-stiffness tensor was defined elementwise as
\begin{equation}
\mathbf D_s=\frac{E_{\mathrm{post}}h}{1-\nu^2}\left[(1-f_a)\mathbf I+2f_a\left(0.65\,\mathbf e_\theta\otimes\mathbf e_\theta+0.35\,\mathbf e_r\otimes\mathbf e_r\right)\right],
\end{equation}
where $f_a=0.34$ is a phenomenological fiber fraction, $\mathbf e_\theta$ is the circumferential direction and $\mathbf e_r$ is the radial direction. This tensor does not claim to be a complete collagen constitutive law; it introduces directional reinforcement into the reduced shell solver.

The load and support were also spatially heterogeneous. The pressure-derived load was amplified in the weakened/thinned cone region to mimic the increased deformation tendency of an ectatic shell,
\begin{equation}
q(x,y)=q_0\left[1+2.15G_{\mathrm{cone}}(x,y)\right]\left(\frac{h_0}{h(x,y)}\right)^2,
\end{equation}
while the support term was reduced near the cone. The outer limbal rim was fixed. This choice intentionally favors comparison of treatment fields under a common geometry rather than prediction of a patient-specific postoperative shape.

\subsection{Optical readout and Kmax-equivalent index}
The deformed anterior surface was reconstructed as a reference spherical cap plus ectatic shape component plus pressure displacement. Zernike coefficients were fitted over a 6-mm pupil using the anterior-surface optical path difference. Higher-order RMS was computed after removing piston, tilt, defocus and astigmatism.

Because a reduced scalar shell model does not contain a full anterior/posterior ray-tracing calculation, we report a Kmax-equivalent severity index rather than a clinical keratometric Kmax. It was calibrated so that the untreated numerical eye corresponds to a moderate keratoconus-like value of 52.50 D:
\begin{equation}
K_{\mathrm{eq}}=43.5+9.0\left[0.72\left(\frac{d_{\mathrm{cone}}}{d_{\mathrm{cone},0}}\right)+0.28\left(\frac{\Psi}{\Psi_0}\right)\right].
\end{equation}
This index is used only for within-model comparison of treatment patterns. It should not be interpreted as a patient-specific topographic measurement. Similarly, the Zernike coefficients reported below are reduced-model optical readouts obtained from the reconstructed anterior surface; they should not be compared directly with clinical aberrometry without full anterior--posterior ray tracing and patient-specific alignment.

\section{Results}
\subsection{Biomechanical--optical metrics}
The model produced a reproducible comparison across treatment masks (Table~\ref{tab:results}). As expected, the untreated case showed the largest cone displacement, vertical coma and HOA RMS. Uniform stiffening was effective mechanically: among the simple masks, it produced the largest reduction in cone displacement. It did not, however, eliminate the asymmetric optical component. Cone-sector and coma-gradient stiffening reduced vertical coma more strongly than uniform CXL, but each introduced its own displacement and strain-energy trade-off. The inverse-smooth mask gave the most balanced response, combining a Kmax-equivalent reduction similar to broad stiffening with lower HOA RMS and a smoother dose distribution.

\begin{table}[t]
\centering
\caption{Enhanced reduced FEM and optical readout for treatment masks. $K_{\rm eq}$ is a calibrated Kmax-equivalent severity index for within-model comparison, not a clinical keratometric measurement.}
\label{tab:results}
\begin{tabular}{lrrrrr}
\toprule
Case & $K_{\rm eq}$ (D) & Cone disp. ($\mu$m) & Vertical coma ($\mu$m) & HOA RMS ($\mu$m) & Strain-energy norm. \\
\midrule
Untreated & 52.50 & 280.2 & 13.413 & 6.571 & 1.000 \\
Uniform & 48.39 & 142.3 & 6.254 & 3.208 & 0.637 \\
Cone sector & 50.51 & 198.0 & 4.765 & 3.071 & 0.966 \\
Partial annulus & 50.60 & 235.6 & 4.954 & 2.606 & 0.654 \\
Coma-gradient & 49.28 & 168.9 & 4.676 & 2.996 & 0.742 \\
Cone Gaussian & 51.25 & 250.2 & 7.967 & 3.700 & 0.780 \\
Inverse smooth & 48.81 & 154.6 & 5.222 & 2.965 & 0.690 \\
\bottomrule
\end{tabular}
\end{table}

\begin{figure}[t]
\centering
\includegraphics[width=0.85\linewidth]{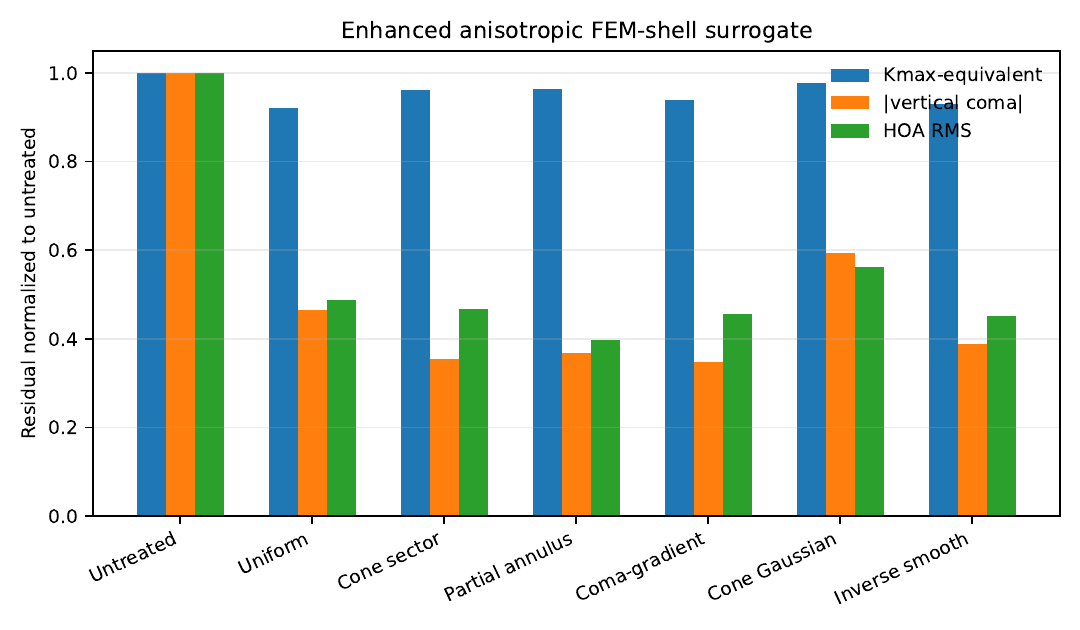}
\caption{Residual biomechanical--optical metrics normalized to the untreated keratoconus-like model. Sectorial and gradient masks can target coma, but the inverse-smooth mask provides a more balanced response across Kmax-equivalent severity, vertical coma and HOA RMS.}
\label{fig:metrics}
\end{figure}

Figure~\ref{fig:metrics} summarizes residual metrics normalized to the untreated case. The main message is not that one sector is universally superior. Different masks optimize different quantities. In this geometry, the cone-sector mask produced the strongest vertical-coma reduction, whereas uniform CXL reduced displacement most efficiently. The inverse-smooth mask avoided the most extreme trade-offs by combining broad stabilization with angularly graded correction.

\subsection{Spatial maps}
\begin{figure}[H]
\centering
\includegraphics[width=0.98\linewidth]{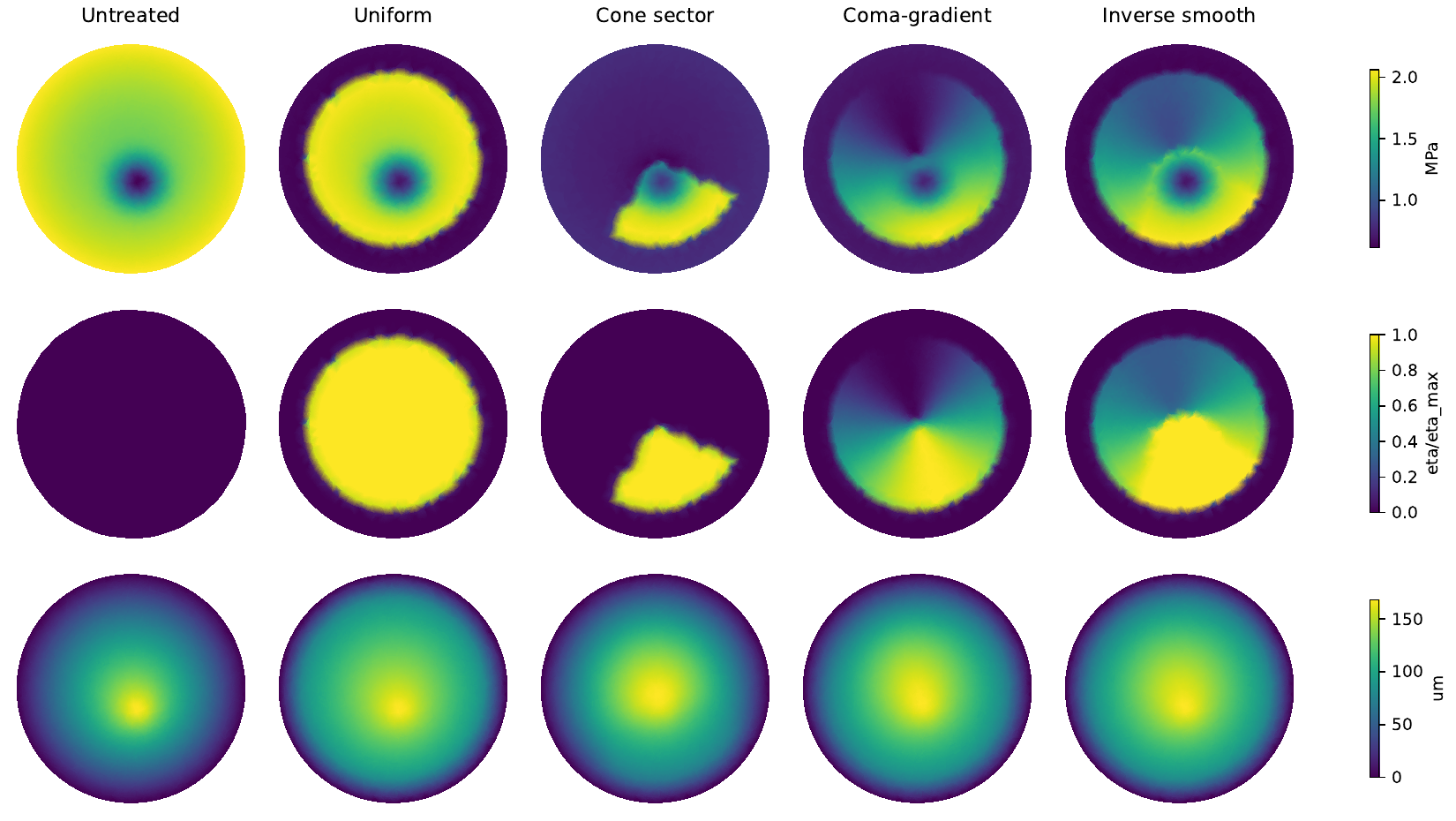}
\caption{Enhanced reduced FEM maps. Top row: post-treatment effective modulus. Middle row: normalized stiffening mask. Bottom row: pressure-induced anterior displacement under 15 mmHg. The inverse-smooth design avoids a sharp sector boundary while preserving cone-directed reinforcement.}
\label{fig:maps}
\end{figure}

Figure~\ref{fig:maps} shows effective modulus, normalized treatment mask and pressure-induced displacement for representative cases. The untreated cornea shows the largest displacement in the weakened inferotemporal region. Uniform CXL suppresses displacement by raising stiffness over a broad central area. The cone-sector mask localizes treatment but introduces sharper angular gradients. The inverse-smooth mask distributes dose more gradually, which is desirable because abrupt stiffness transitions can create new stress concentrations.

\begin{figure}[H]
\centering
\includegraphics[width=0.92\linewidth]{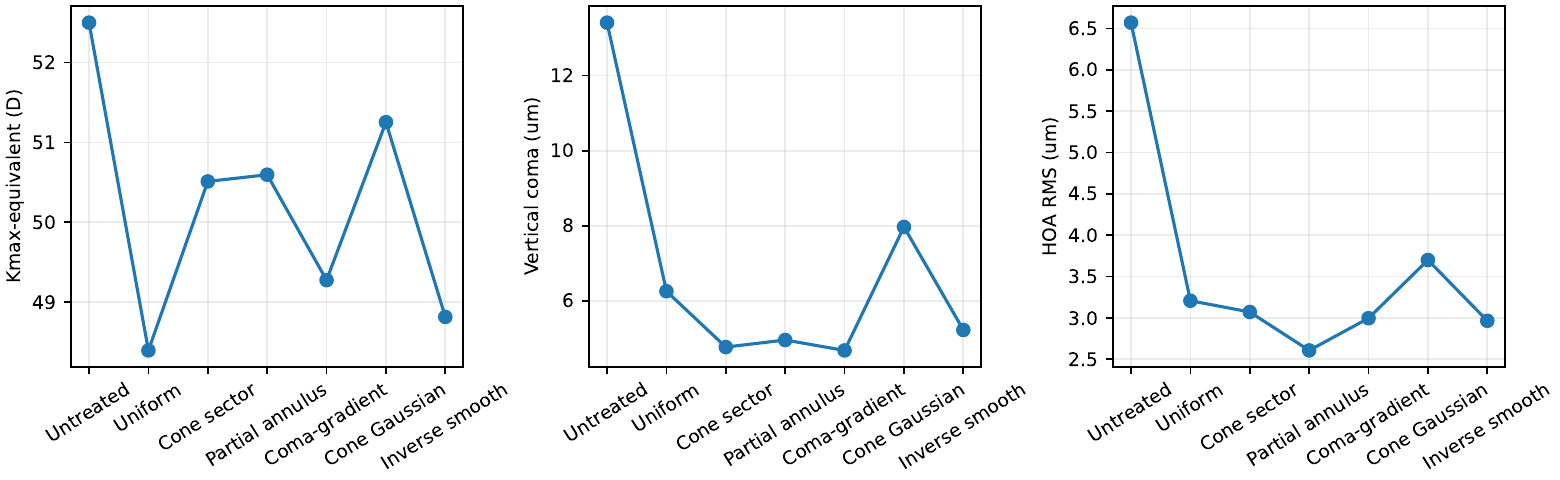}
\caption{Optical metrics from the deformed anterior surface. The Kmax-equivalent index, vertical coma and HOA RMS respond differently to the treatment masks, showing that customized CXL requires multi-objective optimization.}
\label{fig:optical}
\end{figure}

Figure~\ref{fig:optical} reports the optical metrics extracted from the deformed surface. The key observation is that optical improvement and mechanical stabilization are related but not identical. A mask can reduce displacement while leaving residual coma; another can reduce coma while leaving a larger Kmax-equivalent severity. This supports a combined objective rather than a single topographic target.

\begin{figure}[H]
\centering
\includegraphics[width=0.95\linewidth]{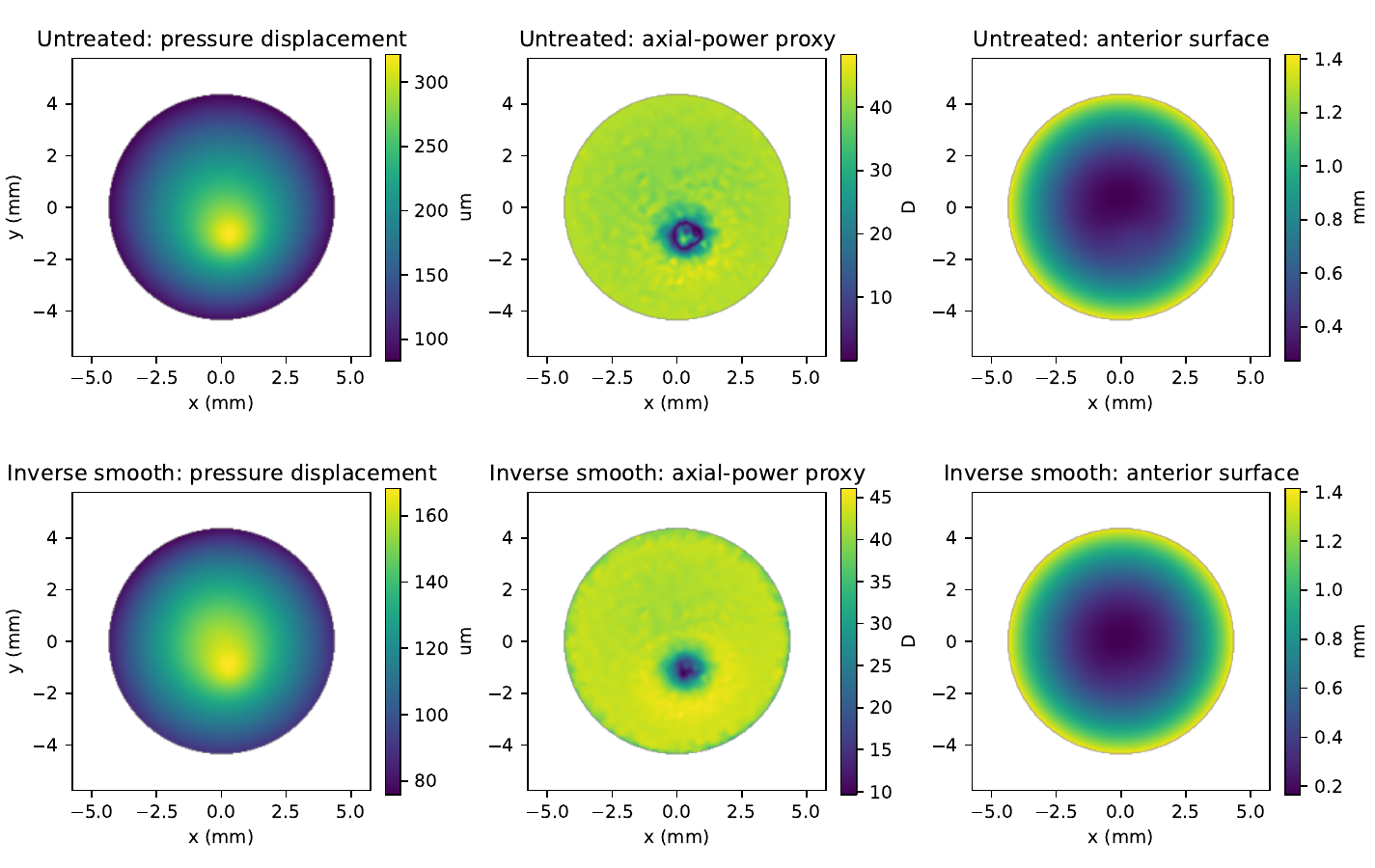}
\caption{Untreated versus inverse-smooth surface response. The optimized smooth mask reduces the displacement field and regularizes the axial-power proxy in the reduced model.}
\label{fig:surface}
\end{figure}

Figure~\ref{fig:surface} compares untreated and inverse-smooth surface responses. The inverse-designed mask reduces the pressure displacement amplitude and regularizes the axial-power proxy. These maps are model outputs, not clinical tomography measurements.

\subsection{IOP sensitivity}
\begin{figure}[H]
\centering
\includegraphics[width=0.92\linewidth]{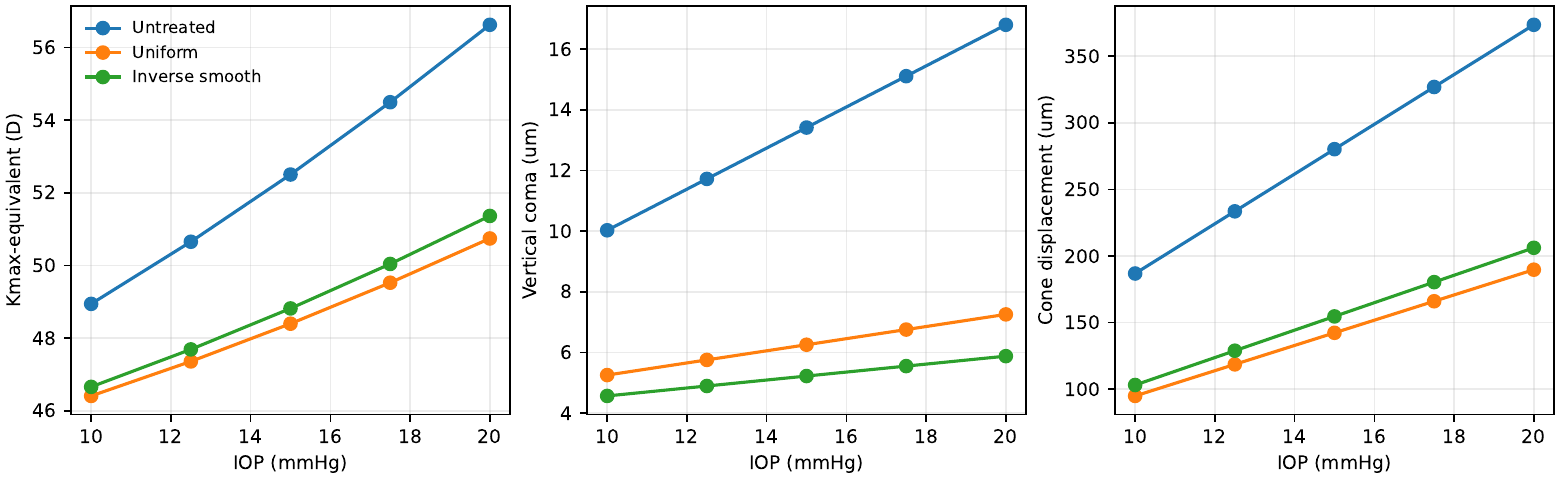}
\caption{IOP sensitivity for untreated, uniform and inverse-smooth cases. The plotted severity index is Kmax-equivalent, not clinical keratometry. Crosslinked cases reduce pressure-induced deformation relative to the untreated model over the simulated pressure range.}
\label{fig:iop}
\end{figure}

The model was also evaluated at 10, 12.5, 15, 17.5 and 20 mmHg for untreated, uniform and inverse-smooth cases (Fig.~\ref{fig:iop}). Increasing pressure increased displacement and coma in all cases, but the crosslinked cases were less pressure-sensitive than the untreated model. This test is important because CXL is a biomechanical intervention: a treatment pattern should be judged not only by its nominal shape at one pressure, but also by whether it reduces deformation under physiological loading.

\FloatBarrier

\section{Discussion}
The central result is that customized CXL is better treated as a spatial control problem than as a simple decision to stiffen the steepest corneal region. Uniform CXL asks whether broad stiffening can reduce deformation. Sectorial and inverse-designed CXL ask a more specific question: which stiffness field best restores mechanical balance while improving the optical readout?

The simulations support three practical points. First, localization is not automatically beneficial. A sharp cone-sector mask can reduce vertical coma in this particular geometry, but it also creates stronger spatial gradients and does not minimize every residual metric. Second, broad stiffening is mechanically robust: it efficiently reduces pressure-induced displacement, but it is not necessarily coma-optimal. Third, a smooth inverse-designed mask can compromise between stabilization, coma reduction, HOA reduction and dose smoothness. This trade-off is the main reason to formulate the treatment as an optimization problem rather than to select a mask only from topography.

These findings also help interpret the growing clinical and experimental literature on localized CXL. Cone-centered treatment can stabilize selected progressive cases and may improve Kmax or visual acuity in a subset of eyes \citep{VorobichikBerar2025}, but the response depends on the spatial relation between the treatment zone, the cone, the thickness map and the mechanical boundary conditions. Brillouin studies of localized CXL have also shown that the biomechanical effect is not confined to the irradiated zone; a transition region forms around the treated area \citep{Webb2019}. This supports the use of smoothness penalties in the inverse design rather than sharp, binary treatment edges.

These findings are consistent with a mechanobiological view of keratoconus. The disease is not merely an anterior-surface shape abnormality; it is a localized loss of load-bearing capacity in an anisotropic, hydrated and pressurized shell. Continuum models are efficient and clinically practical, whereas micromechanical models are better suited for collagen architecture, crosslinks and tissue thinning \citep{DeBellisPandolfi2024,Kory2024}. The present model sits between those levels. It does not resolve individual collagen lamellae, but it does include spatial stiffness loss, local thinning and anisotropic reinforcement, which are the minimum ingredients needed to compare candidate stiffening fields. A stricter microstructural interpretation would replace the scalar stiffening field by changes in collagen-fiber families, proteoglycan matrix response and crosslink architecture; that is the natural next step toward the type of multiscale modeling advocated in recent corneal mechanics work.

Genipin is used here as a modeling agent because its delivery can be expressed through chemical transport and dose-response kinetics. Ex vivo and animal studies suggest that genipin can stiffen ocular collagenous tissues \citep{Gharaibeh2018,Tang2019}; nevertheless, translation to human keratoconus would require extensive safety validation. Endothelial toxicity, stromal haze, pigmentation, epithelial permeability, concentration control and long-term tissue response would all have to be addressed before any clinical use. Therefore, the contribution of this work is not a genipin treatment protocol. The contribution is a design framework that could also be applied to riboflavin--UVA fluence patterns, oxygen-modulated CXL, pulsed irradiation, transepithelial approaches or drug-eluting contact lenses.

One possible extension is hybrid mechanical--biochemical reshaping. A contact lens or orthokeratology device could impose a controlled mechanical boundary condition while a low-dose crosslinking agent stabilizes selected components of the deformation state. That idea remains speculative, but it follows naturally from the inverse-design viewpoint: first define the mechanical and optical target, then design the spatial stiffness change needed to approach it.

\section{Limitations}
This study is theoretical and computational. It should not be interpreted as a validated clinical protocol. The numerical values in Tables~\ref{tab:inputs} and \ref{tab:results} are outputs of a simplified corneal shell surrogate, not patient-derived outcomes. The Kmax-equivalent index is included only for within-model comparison and must not be interpreted as clinical keratometry.

The solver does not include full three-dimensional hyperelasticity, lamellar collagen dispersion, epithelial remodeling, hydration, wound healing, oxygen kinetics, riboflavin photochemistry or endothelial safety modeling. The effective modulus used here is therefore not a calibrated constitutive parameter; it is a reduced stiffness field used for controlled comparison of masks. The inverse problem is also non-unique: different stiffness fields may produce similar anterior shapes, and an optical improvement may not be the mechanically safest solution. For that reason, safety constraints should be primary. A clinically acceptable treatment would need to avoid excessive endothelial exposure, abrupt stiffness gradients, destabilizing stress concentrations and unnecessary treatment of thin tissue. For this reason, the present conclusions should be read as design hypotheses to be tested with patient-specific geometry, calibrated constitutive laws and independent biomechanical validation rather than as predicted clinical outcomes.

\section{Conclusions}
Sectorial customized CXL can be formulated as an inverse biomechanical design problem for keratoconus. The target is not simply to increase corneal stiffness, but to select a spatial stiffness distribution that redistributes stress, reduces cone-driven deformation and improves optical quality. In the present anisotropic reduced shell FEM surrogate, uniform, sectorial, annular, gradient and inverse-designed masks produced distinct trade-offs. Smooth inverse-designed masks were the most conservative design principle because they combined broad stabilization with coma-directed correction while avoiding abrupt treatment boundaries. Genipin-mediated stiffness modulation provides a useful model system for studying depth-dependent and angularly patterned biochemical reinforcement, but experimental calibration and more complete three-dimensional modeling are required before clinical translation.

\section*{Declarations}
\textbf{Funding.} No external funding was received for this theoretical study.\\
\textbf{Competing interests.} The authors declare no competing interests.\\
\textbf{Ethics approval.} Not applicable. This work is theoretical and uses no human or animal subjects.\\
\textbf{Consent to participate.} Not applicable.\\
\textbf{Consent for publication.} Not applicable.\\
\textbf{Data availability.} The reduced finite-element Python script, input-parameter table, simulation outputs, IOP-sensitivity table and figure-generation files are included with the source package.\\
\textbf{Author contributions.} J.S.-M. conceived the study, developed the theoretical framework, prepared the mathematical model and wrote the manuscript. A.A.-T. contributed clinical ophthalmology context, interpretation of the keratoconus and customized crosslinking framework, and manuscript review.

\printbibliography
\end{document}